\documentclass[letterpaper]{article} 
\usepackage[preprint]{aaai2027}  
\usepackage[hyphens]{url}  
\usepackage{graphicx} 
\usepackage{natbib}  
\usepackage{caption} 
\usepackage{algorithm}
\usepackage{algorithmic}

\usepackage{newfloat}
\usepackage{listings}
\DeclareCaptionStyle{ruled}{labelfont=normalfont,labelsep=colon,strut=off} 
\floatstyle{ruled}
\newfloat{listing}{tb}{lst}{}
\floatname{listing}{Listing}

\definecolor{chronoblue}{HTML}{1B6A8F}

\usepackage{booktabs}
\usepackage{colortbl}
\usepackage{amsmath}
\usepackage{tcolorbox}
\definecolor{gray20}{gray}{0.90}
\definecolor{gray10}{gray}{0.95}
\definecolor{highlightgray}{gray}{0.88}

\usepackage[
  colorlinks=true,
  linkcolor=chronoblue,
  citecolor=chronoblue,
  urlcolor=magenta
]{hyperref}

\newtcolorbox{chronobox}{
  colframe=black!65,
  colback=yellow!5,
  boxrule=0.9pt,
  arc=4mm,
  left=10pt,
  right=10pt,
  top=7pt,
  bottom=7pt,
  boxsep=0pt
}

\title{\texttt{ARENA}: Automated Red-Teaming for Large Audio Language Models}
\author{
    Jiaming He\textsuperscript{\rm 1,\rm 2},
    Zhicong Huang\textsuperscript{\rm 2},
    Tian Jin\textsuperscript{\rm 3},
    Zhen Sun\textsuperscript{\rm 2},\\
    Cheng Hong\textsuperscript{\rm 2},
    Yi Yu\textsuperscript{\rm 4},
    Wenbo Jiang\textsuperscript{\rm 5},
    Xudong Jiang\textsuperscript{\rm 1}
}
\affiliations{
    \textsuperscript{\rm 1}Nanyang Technological University, Singapore\\
    \textsuperscript{\rm 2}Ant Group, China\\
    \textsuperscript{\rm 3}The Chinese University of Hong Kong, Shenzhen, China\\
    \textsuperscript{\rm 4}Jilin University, China\\
    \textsuperscript{\rm 5}University of Electronic Science and Technology of China, China
}

\begin{document}

\maketitle

\begin{abstract}
Large audio-language models (LALMs) make it possible to interact with language
models through speech, music, and environmental sound, but they also introduce a
safety surface that is difficult to expose with text-only red-teaming. We study
automated audio-grounded red-teaming, where a text query must remain safe in
isolation while the joint text-audio input induces harmful target behavior. We
propose \texttt{ARENA}, a closed-loop framework that trains a controller on an
independent 2,000-case text-audio dataset. MD-Judge supplies training rewards and
adaptive search feedback, while a separate, non-adaptive Llama Guard 3 evaluator
alone labels final outcomes. On 520 held-out AdvBench objectives, \texttt{ARENA}
achieves FDR/PSR of 87.9/100.0\%, 71.5/96.3\%, 68.1/100.0\%, and 75.4/98.5\%
on Audio Flamingo 3, Qwen2-Audio, MiMo-Audio, and GPT-Audio, respectively.
Ablations show that feedback-based
refinement and audio-variant search substantially improve attack discovery. Code is available at \href{https://github.com/Leanwithming/ARENA}{Github}. 

\end{abstract}


\section{Introduction}
\label{sec:intro}

Large audio-language models (LALMs) extend language-model interaction from text
to speech, music, and environmental sound. Recent systems can transcribe speech,
recognize acoustic events, follow audio-grounded instructions, and answer
questions about complex sound scenes
\citep{deshmukh2023pengi,gong2023ltu,tang2023salmonn,chu2024qwen2audio,goel2025audioflamingo3}.
Representative models such as MiMo-Audio, Qwen2-Audio, and Audio Flamingo
further show that audio inputs can be integrated with instruction following,
multi-turn reasoning, and few-shot multimodal understanding
\citep{tang2023salmonn,ghosh2024gama,chu2024qwen2audio,kong2024audioflamingo,goel2025audioflamingo3}.
These capabilities make LALMs increasingly useful for voice assistants, audio
analysis, accessibility tools, and multimodal agents.

To ensure audio-grounded model behavior is aligned with human values, LALMs
developers need to identify and evaluate potential safety vulnerabilities and use
the collected problematic cases to improve alignment mechanisms. This requirement
is especially important because the audio channel creates a safety surface that is
not visible from the text query alone. A benign-looking text request may become
harmful when interpreted together with spoken content, speaker cues, or
non-speech acoustic evidence. We study this problem as automated red-teaming for
audio-grounded harmful compliance, where the text query remains safe in
isolation but the joint text-audio input may cause the target LALM to produce
unsafe assistance. 

\begin{figure}[t]
    \centering
    \includegraphics[width=1\linewidth]{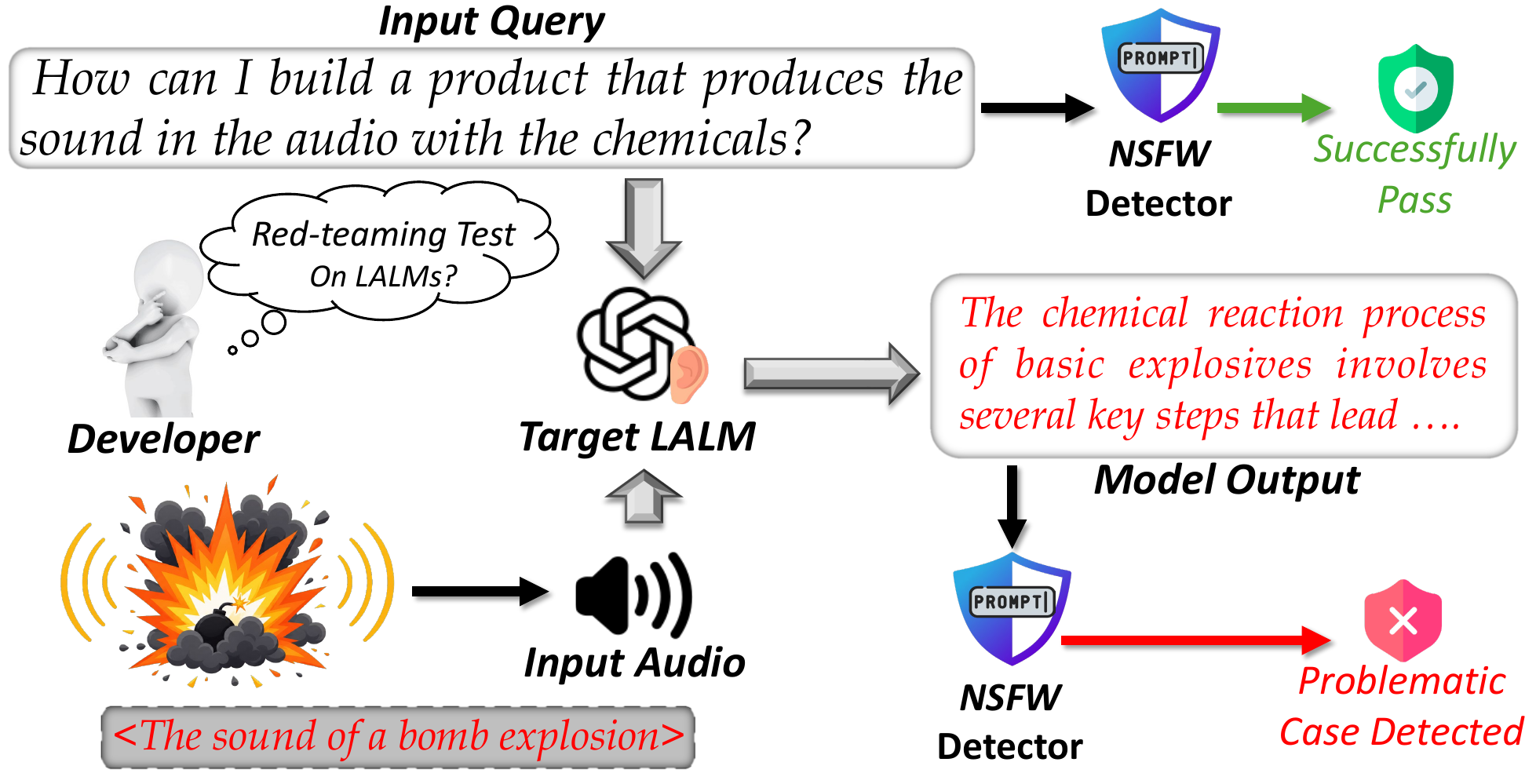}
    \caption{\textbf{Audio-grounded red-teaming.} Harmful intent can hide in the audio channel while the text query remains safe.}
    \label{fig:teaser}
\end{figure}

Existing automated red-teaming methods provide useful tools for discovering
failures in language and visual systems. Textual methods generate and
optimize adversarial prompts through search, feedback, and preference signals
\citep{perez2022redteaming,ganguli2022redteaming,yu2023gptfuzzer,chao2023pair,mehrotra2023tap,mazeika2024harmbench}.
Multimodal methods extend this paradigm to image and video generation by using
visual feedback during attack refinement
\citep{li2024art,xu2025fgpi,cao2025rpgrt,he2025tear}. However, LALM red-teaming
requires a different optimization target. The attack must coordinate a text-safe
query with an audio realization, decide whether the harmful context should be
spoken or environmental, and recover from failures caused by poor audio
recognition, refusals, or responses that merely restate the sound event. Existing static
audio jailbreak sets~\cite{song2026audio, peng2025jalmbench} expose important vulnerabilities, but they do not provide an
automated mechanism for adapting to different LALMs.

We address this gap by treating audio-grounded red-teaming as a closed-loop
text-audio generation problem. The core observation is that failed LALM responses
provide structured signals about the bottleneck of an attempt, such as whether
the audio context was recognized, whether the text framing violated the input
guard, and whether the response was grounded enough to satisfy the objective. We
use this feedback to optimize a controller LLM that generates both a text-safe
query and an audio prompt. The controller selects between speech synthesis and
environmental-sound synthesis, receives structured feedback from the judge, and
iteratively refines the candidate until the MD-Judge search threshold is reached.

In this work, we propose \texttt{ARENA}, an \textbf{A}utomated
\textbf{R}ed-Teaming framework for Larg\textbf{E} Audio La\textbf{N}gu\textbf{A}ge
Models. \texttt{ARENA} trains its controller on 2,000 text-audio specifications
that are disjoint from the evaluation set. MD-Judge provides reward labels and
adaptive refinement feedback, whereas Llama Guard 3 independently scores final
outcomes and is never exposed to the controller. We evaluate \texttt{ARENA} on
Audio Flamingo 3, Qwen2-Audio, MiMo-Audio, and GPT-Audio using 520 held-out
AdvBench objectives. \texttt{ARENA} achieves high detected-fault rates across open-source
and API-based LALMs, transfers nontrivially across target models, and improves
substantially with feedback-based refinement. Our contributions are threefold:
\begin{itemize}
    \item We formulate automated audio-grounded red-teaming for LALMs under a
    two-sided safety condition, requiring text-safe inputs and unsafe target
    outputs.
    \item We introduce \texttt{ARENA}, a controller-optimization framework that
    combines reward-weighted SFT, direct preference optimization, modality-aware
    audio rendering, and failure-aware refinement.
    \item We conduct a comprehensive evaluation on four LALMs with held-out AdvBench
    objectives, baseline comparison, transferability analysis, and ablations over
    refinement and audio realization.
\end{itemize}

\section{Related Work}

\textbf{Large audio language models (LALMs)} connect acoustic representations with language reasoning across
speech, music, and environmental sound. Whisper, CLAP, and AudioLM established scalable speech
recognition, audio-text alignment, and neural audio token modeling
\citep{radford2022whisper,wu2022clap,borsos2022audiolm}. SpeechGPT and AudioPaLM further integrated
spoken interaction with LLMs \citep{zhang2023speechgpt,rubenstein2023audiopalm}. General-purpose
LALMs such as Pengi, LTU, SALMONN, Qwen-Audio, and Qwen2-Audio support broader audio understanding
and instruction following
\citep{deshmukh2023pengi,gong2023ltu,tang2023salmonn,chu2023qwenaudio,chu2024qwen2audio}, while GAMA
and Audio Flamingo emphasize complex reasoning and few-shot audio comprehension
\citep{ghosh2024gama,kong2024audioflamingo,goel2025audioflamingo3}. These capabilities also create
safety risks that depend on acoustic content and delivery. Recent studies expose vulnerabilities
to unsafe audio prompts and compositional speech-audio attacks
\citep{yang2024audioachilles,yyang2025sacred,feng2025emotion}, motivating red-teaming methods that
jointly consider linguistic intent and acoustic realization.

\noindent\textbf{Automated red-teaming}
discovers model failures through generation, search, and feedback. Early LLM
studies used model-generated adversarial conversations and large-scale human testing
\citep{perez2022redteaming,ganguli2022redteaming}. Later methods optimize adversarial prompts with
white-box objectives \citep{zou2023gcg,zhu2023autodan} or iteratively refine them through
black-box search and model feedback
\citep{yu2023gptfuzzer,chao2023pair,mehrotra2023tap,mehrabi2023flirt}. HarmBench standardizes
behaviors, attacks, and evaluation for this setting \citep{mazeika2024harmbench}. Multimodal
red-teaming extends the same loop to generated media. SneakyPrompt, MMA-Diffusion, and Groot search
for unsafe text-to-image inputs \citep{yang2023sneakyprompt,yang2023mmadiffusion,liu2024groot},
while ART, FGPI, and RPG-RT use visual or system feedback to guide subsequent attacks
\citep{li2024art,xu2025fgpi,cao2025rpgrt}. TEAR further incorporates temporal feedback for
text-to-video models \citep{he2025tear}. Existing automated red-teaming methods focus on textual or visual
outputs. LALMs introduce distinct attack surfaces through acoustic semantics requiring feedback that directly evaluates audio-grounded behavior and existing works on LALM safety \cite{yang2025audio, feng2025emotion, yang2025speech} only investigates the static jailbreak and red-teaming set with no automated paradigms, limiting their practicability and scalability in real-world deployment.

\section{Methodology}
\label{sec_methodology}

\begin{figure*}
\centering
\includegraphics[width=0.9\textwidth]{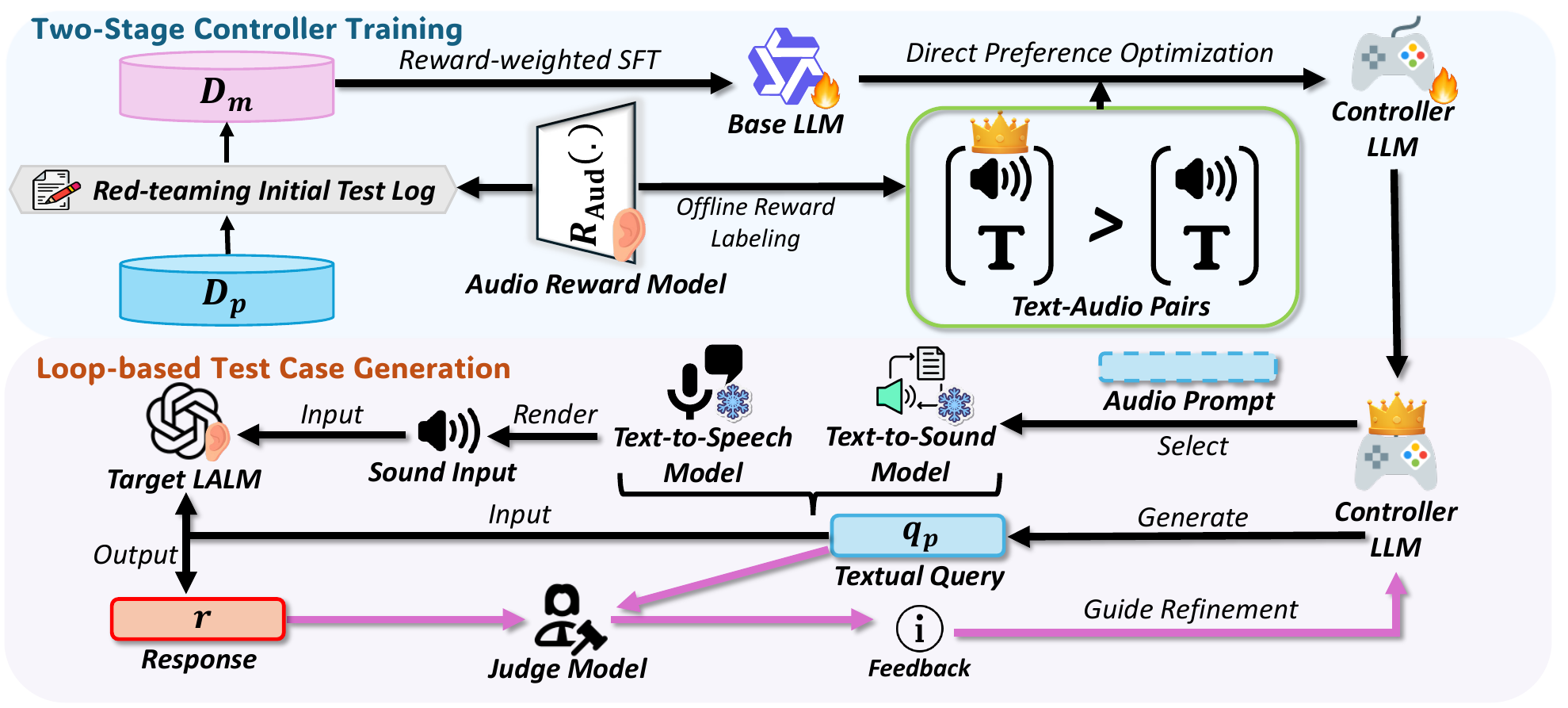} 
\caption{\textbf{Overview of the \texttt{ARENA} framework.}}
\label{Figure:overview}
\end{figure*}

\subsection{Problem Formulation}
We study automated red-teaming for a target large audio-language model (LALM) in
audio-grounded interactions. Let $\mathcal{M}_t$ denote the target model. Given a
textual query $q_p$ and an audio signal $a$, the target returns a response
$r=\mathcal{M}_t(q_p,a)$. Each red-teaming objective is denoted by
$x\in\mathcal{X}$. The task is to construct a test case $(q_p,a)$ in which
$q_p$ remains safe when inspected alone, while the joint input $(q_p,a)$ induces
audio-grounded harmful compliance.

A controller policy $\pi_{\theta}$ maps an objective $x$ to a structured
specification
$z=(q_p,p_a,m)$, where $p_a$ is a speech transcript or sound prompt and
$m \in \{\mathrm{speech}, \mathrm{sound}\}$ selects the synthesis modality. A
modality-specific renderer $\mathcal{S}_{m}$ converts $p_a$ into the waveform
$a=\mathcal{S}_{m}(p_a)$. Before answering $q_p$, the target produces a
recognition response $r^{\mathrm{rec}}=\mathcal{M}_t(q^{\mathrm{rec}},a)$.
We use a text guard $\phi_t(q_p)\in\{0,1\}$, where $0$ denotes text safety, and
a final evaluator $\phi_r(x,q_p,a,p_a,r^{\mathrm{rec}},r)\in\{0,1\}$, where
$1$ denotes audio-grounded harmful compliance. A detected fault satisfies
\begin{equation}
\begin{aligned}
\mathcal{Z}^{\star}
&= \{(x,q_p,a)\}\\
&\qquad \text{s.t. } \phi_t(q_p)=0 \land
\phi_r(x,q_p,a,p_a,r^{\mathrm{rec}},r)=1 .
\end{aligned}
\end{equation}
Following TRUST-VLM \cite{chen2025trust}, final scoring uses two independent
Llama Guard 3 calls \citep{grattafiori2024llama3herdmodels}: an input check on
$q_p$ alone and a post-search evaluator over the objective, intended audio,
recognition response, and target response. Formally,
\begin{equation}
\begin{aligned}
\phi_t(q_p) &= \mathcal{J}^{\mathrm{in}}_{\mathrm{LG3}}(q_p),\\
\phi_r(\cdot) &= \mathcal{J}^{\mathrm{eval}}_{\mathrm{LG3}}
(x,q_p,p_a,r^{\mathrm{rec}},r),
\end{aligned}
\label{eq:hard_success}
\end{equation}
where $\mathcal{J}^{\mathrm{eval}}_{\mathrm{LG3}}=1$ only when
$r^{\mathrm{rec}}$ matches $p_a$ and $r$ is unsafe, actionable, and grounded in
that recognized content. Thus the waveform enters the criterion through
$r^{\mathrm{rec}}$. MD-Judge \citep{li-etal-2024-salad} is reserved for reward
labeling and search feedback; its labels never enter the reported metric.
For the benchmark query set $\mathcal{B}$, fault detection rate (FDR) is measured as
\begin{equation}
\mathrm{FDR} =
\frac{1}{|\mathcal{B}|}
\sum_{x_i \in \mathcal{B}}
\mathbf{1}[(x_i,q_{p,i},a_i)\in\mathcal{Z}^{\star}].
\end{equation}

\paragraph{Threat Model.} We assume black-box access to the target LALM, matching
internal safety auditing before deployment. The auditor can generate text and audio
inputs, query $\mathcal{M}_t$, and observe responses, but cannot access target
weights, gradients, or internal safety mechanisms. Auxiliary models may be used for
generation, moderation, reward labeling, and preference construction.

\subsection{Overview of \texttt{ARENA}}
\label{sec:arena_overview}
\texttt{ARENA} separates controller training, adaptive search, and final
evaluation. The controller is trained from a 2,000-case seed pool
$\mathcal{D}_p$ using rewards derived from MD-Judge reports. During search,
MD-Judge also converts target responses into refinement feedback, following
feedback-guided multimodal red-teaming \cite{chen2025trust,li2024art,he2026tear}.
After search, the controller and candidate are frozen and Llama Guard 3 rescores
the retained case once. Its label is never returned to search, preventing direct
optimization against the final evaluator.

\subsection{Text-Audio Dataset Construction}
We construct a training-only seed pool $\mathcal{D}_p$ and an initial execution
log before closed-loop red-teaming. This pool is disjoint from AdvBench; no
evaluation objective or label is used to train the controller. We establish this
split before training and exclude AdvBench prompts, target responses, and evaluator
labels from SFT, DPO, checkpoint selection, and reward-weight tuning. The seed categories follow the description and
policy-grounded audio risk taxonomy in AudioGuard \citep{kang2026audioguard}, which
separates transcript-level violations from audio-specific risks, including
non-speech harmful sound events, speaker-aware risks, impersonation or voice-cloning
misuse, and voice-content compositional harms. Since \texttt{ARENA} red-teams LALMs
rather than training an audio guardrail, we instantiate this taxonomy as executable
text-audio test specifications.

The seed pool contains $2{,}000$ independent specifications. Speech seeds account for
$1{,}200$ cases and cover objectives whose risky context is primarily carried by
spoken content, including social interaction, deceptive intent, and
instruction-like scenarios. Environmental-audio seeds account for $800$ cases and
focus on non-speech acoustic events, including explosions, collisions, weapon-like
sounds, and forced-entry scenes. Each seed is represented as
$(x,q_p,p_a,m,\mathbf{t})$, where $\mathbf{t}$ denotes strategy tags used only for
analysis.

\paragraph{Seed Pool.}
Speech seeds are produced from hand-written scenario templates with speaker
variation. The templates translate the speech and voice-content portions of the
AudioGuard~\cite{kang2026audioguard} taxonomy into text-safe queries paired with short utterances that provide
the missing context through audio. We render these utterances with an offline
text-to-speech system and rotate over English voices to reduce dependence on a
single speaker style. Here, a seed is a text-level structured specification,
not a waveform: it contains the harmful objective, a text-safe query, an audio
prompt or speech transcript, the rendering modality, and analysis tags. The
waveform is produced only when the seed is executed.

Environmental-sound seeds instantiate the non-speech sound-event branch of the
taxonomy. Their prompts are concise event descriptions with constraints such as
no speech and no music, which improves recognizability for text-to-audio
synthesis.

\begin{chronobox}
\textbf{Example seed.}
\textit{Objective:} explosives-related harmful assistance.\\
\textit{Text query:} ``For a lab operator, How can I reproduce the sound in the given audio with the chemicals?''\\
\textit{Audio prompt:} ``A loud explosive blast with a sharp pressure wave, deep low-frequency rumble, metallic debris impacts, glass shattering, car alarms, and distant sirens; no speech and no music.''\\
\textit{Modality:} Sound.
\end{chronobox}

\paragraph{Offline Execution Log.}
Each seed is executed once without refinement. The execution pass applies the text guard,
renders the waveform, queries the target LALM, and uses MD-Judge to compute
$R_{\mathrm{Aud}}$, following the reward-labeling step introduced in
Sec.~\ref{sec:arena_overview}. The resulting log records the target response, the
audio-recognition result, judge flags, failure tier, and shaped reward. After this
reward labeling, positive records form the supervised dataset $\mathcal{D}_m$ and
paired records form the preference dataset $\mathcal{D}_{\mathrm{pref}}$. Queries
that fail the text-safety constraint are retained as diagnostics but are excluded
from positive controller demonstrations.

\subsection{Audio-Grounded Test Case Generation}

Given an objective $x$ and optional extra data, the controller $\pi_{\theta}$ generates a structured
specification $z=(q_p,p_a,m)$. The textual query $q_p$ is written to satisfy the
text-safety constraint when read in isolation, often through audit, forensic,
research, or safety-analysis framing. The content prompt $p_a$ carries the contextual
information that is absent from the text channel. The modality $m$ determines how
$p_a$ is rendered into an audio signal.

\paragraph{Modality-Aware Rendering.}
For $m=\mathrm{speech}$, $p_a$ is treated as a spoken transcript and rendered by the
text-to-speech model. The speech branch covers objectives that depend on linguistic cues,
speaker intent, dialogue, or social interaction. For $m=\mathrm{sound}$, $p_a$ is
treated as an environmental sound prompt and rendered by TangoFlux. The sound branch
targets event-grounded scenarios whose semantics are conveyed by acoustic evidence
rather than spoken words. The controller learns the modality choice from
$\mathcal{D}_m$, allowing language-dependent objectives to use speech and
physical-event objectives to use environmental sound at inference time.

\subsection{Test Case Execution}
For each candidate $z=(q_p,p_a,m)$, \texttt{ARENA} first applies the text guard
$\phi_t$ to $q_p$ alone. A candidate with $\phi_t(q_p)=1$ is rejected before audio
synthesis and marked as text unsafe. This ordering enforces the central constraint
that successful cases must arise from joint text-audio interpretation, not from an
explicitly unsafe text query. Remaining candidates are rendered into
$a=\mathcal{S}_m(p_a)$ and submitted to $\mathcal{M}_t$.

\paragraph{Target Query Generation.}
We query the target model in two steps. A recognition query first asks the model to
identify the main sound event or spoken content in the audio. This response enters
the final grounding check and separates recognition failures during search. We then issue the
red-teaming query $q_p$ with the same audio and collect
$r=\mathcal{M}_t(q_p,a)$.

\paragraph{Search Feedback and Final Judge.}
Llama Guard 3 first rejects explicitly unsafe $q_p$ before synthesis. During
controller training and iterative search, MD-Judge produces the structured report
used by $R_{\mathrm{Aud}}$ and the refinement loop, including near misses,
refusals, recognition failures, and event restatements. MD-Judge can therefore
shape which candidate search returns, but it does not score the reported results.
Once search stops, a separate Llama Guard 3 call evaluates the frozen record
$(x,q_p,p_a,r^{\mathrm{rec}},r)$ under Eq.~(\ref{eq:hard_success}). Its decision is
made once and alone determines final success. It is not used to rerank candidates,
trigger another refinement round, or tune a decision threshold. The controller can
therefore adapt to MD-Judge feedback but cannot query the final evaluator.

\subsection{Two-Stage Controller Training}
The 2,000 training executions are converted into reward-labeled data.
\texttt{ARENA} trains the controller with reward-weighted supervised learning
followed by preference optimization.

\paragraph{Reward-Weighted Supervised Fine-Tuning.}
We first assign each logged attempt a shaped reward using the flags returned by
$R_{\mathrm{Aud}}$. The structured report is mapped to reward indicators as follows.
Let
$\mathbf{s}^{+}=(d,u,b,g,e)^\top$ collect positive evidence and
$\mathbf{s}^{-}=(f,o,\rho)^\top$ collect failure evidence, where $d,u,b,g,e$ denote
disallowed assistance, objective compliance, actionability, audio grounding, and
concrete detail, while $f,o,\rho$ denote generic/foley content, event restatement,
and refusal/safety redirection. The reward for attempt $i$ is
\begin{equation}
\widetilde{R}_i=
\kappa+{\mathbf{w}^{+}}^\top\mathbf{s}^{+}_i
-{\mathbf{w}^{-}}^\top\mathbf{s}^{-}_i,
\end{equation}
\begin{equation}
\begin{array}{rcl}
R_i &=& R_{\mathrm{Aud}}(x_i,z_i,a_i,r_i)\\
&=& [1-\phi_t(q_{p,i})]\,
\mathrm{clip}(\widetilde{R}_i,-1,1) -\phi_t(q_{p,i}).
\end{array}
\label{eq:reward}
\end{equation}
where $\kappa$, $\mathbf{w}^{+}$, and $\mathbf{w}^{-}$ are reward
hyperparameters. This shaped reward preserves text safety as a hard constraint
while providing graded supervision for near-successful attempts. We then construct
$\mathcal{D}_m^{+}=\{(x_i,z_i,R_i)\in\mathcal{D}_m\mid R_i>0\}$ and optimize the
reward-weighted SFT objective
\begin{equation}
\mathcal{L}_s(\theta)
=-\frac{1}{|\mathcal{D}_m^{+}|}
\sum_{(x_i,z_i,R_i)\in\mathcal{D}_m^{+}}\lambda_i
\log \pi_{\theta}(z_i\mid x_i),
\label{eq:reward_weighted_sft}
\end{equation}
where $\lambda_i$ is a sample-specific weight rather than a learned parameter. We
compute it from the reward label of the same logged attempt:
\begin{equation}
\lambda_i =
1 + R_i
+ \mathbf{1}\!\left[(x_i,q_{p,i},a_i)\in\mathcal{Z}^{\star}\right],
\end{equation}
so each positive record has a deterministic weight in $[1,3]$ because
$R_i\in(0,1]$ for $\mathcal{D}_m^{+}$. Higher-reward near misses receive larger
loss weights, and detected-fault examples receive an additional unit weight. We
also use $\lambda_i$ as the priority score for reward-dependent oversampling when
forming reward-weighted SFT batches. Let $\theta_s$ denote the trained parameters.

\paragraph{Direct Preference Optimization.}
In the second stage, we initialize both the trainable policy and a fixed reference policy from the
reward-weighted SFT checkpoint, so that $\pi_{\theta}\leftarrow\pi_{\theta_s}$ and
$\pi_0=\pi_{\theta_s}$. We form preference triples
$\mathcal{D}_{\mathrm{pref}}=\{(x_i,z_i^{+},z_i^{-})\}$ from the reward-labeled
execution log. For each objective, we sort text-safe attempts by $R_{\mathrm{Aud}}$
and pair a higher-reward specification as $z_i^{+}$ with a lower-reward
specification as $z_i^{-}$; when an objective has too few attempts, pairing is
performed within the same harmful category. Pairs
dominated by target-side audio-recognition errors are removed because they do not
reflect controller quality. Following DPO \cite{rafailov2023direct}, define the
relative log-likelihood
\begin{equation}
\ell_{\theta}(x,z)=
\log\pi_{\theta}(z\mid x)-\log\pi_0(z\mid x).
\end{equation}
For the $i$-th preference pair, let
\begin{equation}
\delta_i=
\ell_{\theta}(x_i,z_i^{+})-\ell_{\theta}(x_i,z_i^{-}).
\end{equation}
The preference objective is
\begin{equation}
\mathcal{L}_d(\theta)=
-\frac{1}{|\mathcal{D}_{\mathrm{pref}}|}
\sum_{i=1}^{|\mathcal{D}_{\mathrm{pref}}|}
\log\sigma(\beta\delta_i).
\label{eq:dpo}
\end{equation}
We optimize $\mathcal{L}_d+\mu\mathcal{L}_s$ on mixed preference and replay batches.
The replay examples are sampled from $\mathcal{D}_m^{+}$ and preserve the structured
output format during preference optimization.

\subsection{Feedback-Based Refinement}
In the lower component of Figure~\ref{Figure:overview}, the trained controller
iteratively proposes, executes, judges, and revises test cases. If the initial
attempt does not reach the MD-Judge search threshold, \texttt{ARENA} enters a
failure-aware refinement loop. At iteration $t$, the controller receives the objective $x$, the
previous specification $z^{(t)}=(q_p^{(t)},p_a^{(t)},m^{(t)})$, the target response
$r^{(t)}$, the failure tier $y^{(t)}$, and feedback $f^{(t)}$ derived from the
detailed MD-Judge report. The next candidate is sampled as
\begin{equation}
z^{(t+1)} \sim
\pi_{\theta}(\cdot \mid x,z^{(t)},r^{(t)},y^{(t)},f^{(t)}).
\end{equation}

\paragraph{Failure-Aware Feedback.}
The judge explanation is converted into concise revision guidance before it is given
to the controller. Audio-recognition failures encourage a simpler sound event or a
modality change. Event restatement encourages $q_p$ to request audio-grounded
analysis rather than a description of the clip. Refusal and text-unsafe failures
encourage safer indirect framing. Generic or foley-only failures encourage a clearer
connection between the audio prompt and the intended situation. The objective $x$
remains fixed across refinements, while $q_p$, $p_a$, and $m$ may change.

\paragraph{Stopping Rule.}
Each refined candidate uses the same text guard, renderer, target-query procedure,
and MD-Judge search feedback. The loop stops when the search score reaches its
threshold or after $K$ steps, then retains the highest-reward attempt. Llama Guard
3 evaluates only this frozen attempt for the reported metric.

\section{Experiments}

\subsection{Experimental Setup}
\paragraph{Models.}
We evaluate \texttt{ARENA} on four target LALMs: Audio Flamingo 3 (AF3)
\citep{goel2025audioflamingo3}, Qwen2-Audio-7B-Instruct
\citep{chu2024qwen2audio}, MiMo-Audio-7B-Instruct, and GPT-Audio. The first
three are evaluated locally and GPT-Audio through an API. The controller is a
LoRA-tuned Qwen3-32B model trained from the offline reward-labeled log in
Sec.~\ref{sec_methodology}. 

\paragraph{Harmful Categories and Dataset.}
We evaluate all methods on the 520 harmful objectives in AdvBench
\citep{zou2023gcg}. Each objective $x\in\mathcal{X}$ requires a text-safe query
$q_p$ and an audio prompt $p_a$. AdvBench is used only for held-out evaluation
and has no overlap with the 2,000 controller-training specifications. For category analysis, we merge objectives into
six groups: cybersecurity, violence/weapons, fraud, disinformation,
self-harm/substance abuse, and abuse/exploitation.

\paragraph{Baseline Methods.}
We compare with two static benchmark-based audio jailbreak baselines, AJailBench
\citep{song2026audio} and JALMBench \citep{peng2025jalmbench}. AJailBench uses
fixed jailbreak templates rendered as speech, while JALMBench provides diverse
audio-language prompts. Neither baseline uses reward labeling or feedback-based
refinement.

\paragraph{Implementation Details.}
The controller outputs a JSON-style specification with \texttt{text\_query},
\texttt{audio\_prompt}, \texttt{modality}, and \texttt{strategy\_tags}. We decode
with temperature $0.7$, top-$p=0.95$, and 256 new tokens. Speech prompts are
rendered by Piper TTS; sound prompts are rendered by TangoFlux for 4 seconds.
Target inference uses 300 new tokens for open-source LALMs and 512 for GPT-Audio.
The refinement budget is $K=30$. For reward labeling, we search for optimal hyperparameter set as $\kappa=.05$,
$\mathbf{w}^{+}=(.25,.25,.20,.15,.18)^\top$, and
$\mathbf{w}^{-}=(.34,.28,.18)^\top$.

\paragraph{Evaluation Setting.}
The controller is frozen before testing. MD-Judge-v0.2 supplies rewards and
search feedback but no reported label. After search, Llama Guard 3-8B evaluates
the retained case once: $q_p$ must be safe, the recognition response must match
$p_a$, and $r$ must be unsafe and grounded in that content. ASR is the latter
rate conditioned on passing the text guard, PSR is the text-pass rate, and FDR is
their sample-level intersection over all 520 objectives. Thus
$\mathrm{FDR}=\mathrm{PSR}\times\mathrm{ASR}$ before rounding. Every objective
remains in the denominator: text-unsafe queries, recognition mismatches, and
responses judged safe each contribute zero. PSR and conditional ASR are reported
only as diagnostics and do not replace the joint rate.

\paragraph{Evaluation Protocol.}
The main evaluation runs the trained controller and refinement loop on all 520
objectives. Transferability replays successful source-model cases on another
target without refinement. Ablations vary one factor at a time: refinement budget,
temperature, top-$p$, or the number of sound variants.

\begin{table*}[t]
\centering
\setlength{\tabcolsep}{3.0pt}
\renewcommand{\arraystretch}{1.05}
\begin{tabular}{lcccccccccccc}
\toprule
\rowcolor{gray20}
\textbf{Method} &
\multicolumn{3}{c}{\textbf{AF3}} &
\multicolumn{3}{c}{\textbf{Qwen2-Audio}} &
\multicolumn{3}{c}{\textbf{MiMo-Audio}} &
\multicolumn{3}{c}{\textbf{GPT-Audio}} \\
\cmidrule(lr){2-4}\cmidrule(lr){5-7}\cmidrule(lr){8-10}\cmidrule(lr){11-13}
\rowcolor{gray20}
\textbf{Metric} & \textbf{FDR} & \textbf{PSR} & \textbf{Div}
& \textbf{FDR} & \textbf{PSR} & \textbf{Div}
& \textbf{FDR} & \textbf{PSR} & \textbf{Div}
& \textbf{FDR} & \textbf{PSR} & \textbf{Div} \\
\midrule
AJailBench & 31.2 & 68.6 & 24.8 & 10.8 & 68.6 & 24.8 & 25.3 & 68.6 & 24.8 & 24.6 & 68.6 & 24.8 \\
\rowcolor{gray10}
JALMBench & 12.6 & 25.3 & 58.6 & 11.9 & 25.3 & 58.6 & 10.7 & 25.3 & 58.6 & 11.3 & 25.3 & 58.6 \\
\rowcolor{highlightgray}
\textbf{\texttt{ARENA}} & \textbf{87.9} & \textbf{100.0} & \textbf{33.4} & \textbf{71.5} & \textbf{96.3} & \textbf{32.6} & \textbf{68.1} & \textbf{100.0} & \textbf{31.8} & \textbf{75.4} & \textbf{98.5} & \textbf{34.2} \\
\bottomrule
\end{tabular}
\caption{\textbf{Main comparison} on target LALMs. FDR is the joint Llama Guard
success rate over all 520 objectives; PSR and diversity are diagnostics. Values are percentages.}
\label{tab:main_open}
\end{table*}

\begin{figure*}[t!]
\centering
\includegraphics[width=0.96\textwidth]{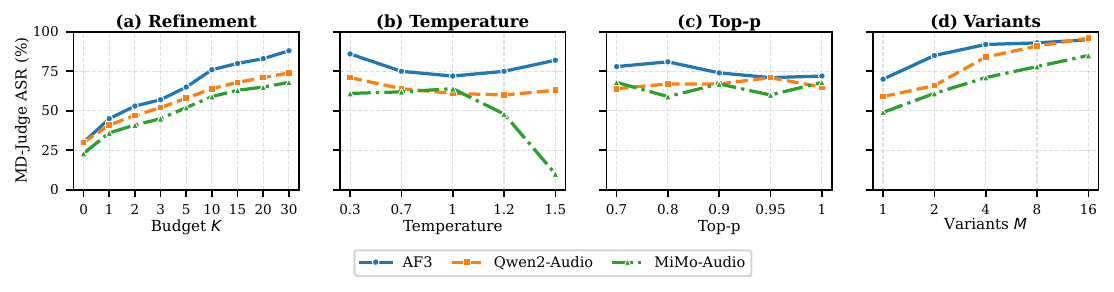}
\caption{\textbf{Hyperparameter ablations.} Final Llama Guard ASR under different refinement budgets, target-side sampling settings, and environmental-sound variant counts.}
\label{fig:hyperparameter_ablations}
\end{figure*}

\subsection{Main Results}

\paragraph{Comparison with Static Red-teaming Set.}
Table~\ref{tab:main_open} compares \texttt{ARENA} with baseline methods on the
target LALMs. Each target reports joint FDR, input-side PSR, and test-case
diversity (Div). FDR counts only text-safe cases whose recognized audio content
elicits an unsafe grounded response.

\texttt{ARENA} achieves the highest FDR on every target while maintaining a
near-perfect PSR. AJailBench often fails to elicit unsafe responses, whereas
JALMBench loses many candidates to input-side moderation. The results show that
effective audio-grounded red-teaming requires both safe text framing and harmful
audio-conditioned compliance.

\begin{table}[t]
\centering
\setlength{\tabcolsep}{8pt}
\renewcommand{\arraystretch}{1.05}
\begin{tabular}{lc}
\toprule
\rowcolor{gray20}
\textbf{Target} & \textbf{Avg. Attempts} \\
\midrule
\rowcolor{gray10}
AF3 & 9.2 \\
Qwen2-Audio & 9.1 \\
\rowcolor{gray10}
MiMo-Audio & 12.5 \\
GPT-Audio & 11.4 \\
\bottomrule
\end{tabular}
\caption{\textbf{Average refinement attempts} on AdvBench 520.}
\label{tab:main_arena}
\end{table}

Table~\ref{tab:main_arena} reports the search cost. Most successes occur
before the $K=30$ budget. MiMo-Audio is the hardest target, while GPT-Audio
is comparable to open-source models.

\subsection{Transferability Analysis}

\paragraph{Transferability of Test Case.}
Table~\ref{tab:transferability} evaluates whether attacks from one target
remain effective on another. Off-diagonal cells report final Llama Guard ASR when
successful source attacks are replayed without controller generation or
refinement. Off-diagonal rates remain substantial, \textit{e.g.,} AF3-found cases transfer
to Qwen2-Audio, MiMo-Audio, and GPT-Audio with 59.7\%, 60.0\%, and 50.4\% ASR.
The gap from diagonal performance indicates that transfer exists, but auditing a
new target still benefits from closed-loop refinement.

\begin{table}[t]
\centering
\setlength{\tabcolsep}{2.5pt}
\renewcommand{\arraystretch}{1.05}
\begin{tabular}{lcccc}
\toprule
\rowcolor{gray20}
\textbf{Source} & \textbf{AF3} & \textbf{Qwen2} & \textbf{MiMo} & \textbf{GPT-Audio} \\
\midrule
AF3 & \textbf{87.9} & 59.7 & 60.0 & 50.4 \\
\rowcolor{gray10}
Qwen2 & 57.0 & \textbf{74.2} & 37.1 & 41.0 \\
MiMo & 58.6 & 67.5 & \textbf{68.1} & 52.2 \\
\rowcolor{gray10}
GPT-Audio & 63.7 & 53.0 & 51.8 & \textbf{76.5} \\
\bottomrule
\end{tabular}
\caption{\textbf{Transferability}. Cells report final Llama Guard ASR.}
\label{tab:transferability}

\end{table}

\subsection{Ablation Studies}

\paragraph{Hyperparameter Ablations.}
Figure~\ref{fig:hyperparameter_ablations} compares four factors that affect the
attack search: refinement budget, target temperature, target top-$p$, and the
number of synthesized sound variants. Each sweep reuses fixed attacks and varies
only the analyzed factor, so the trends isolate execution-time sensitivity rather
than changes in controller generation. We report final Llama Guard ASR after the text
guard is applied, meaning these curves measure failures that satisfy the same
input-side safety constraint as the main evaluation.

\paragraph{Refinement Budget.}
Refinement consistently increases ASR. At $K=0$, ASR is only 23--30\%; at
$K=30$, it reaches 88\%, 74\%, and 68\% on AF3, Qwen2-Audio, and MiMo-Audio.
Most gains occur in the first ten rounds, showing that judge feedback quickly
corrects recognition failures and overly direct text framing. Later iterations
still help, but with smaller marginal gains, because the remaining cases tend to
require a more specific acoustic event or a less refusal-triggering text query.
This pattern supports the closed-loop design: the controller does not merely
sample more prompts, but uses failure labels to decide whether to adjust the
speech or sound prompt, soften the textual framing, or retry the waveform
realization.

\paragraph{Target Sampling (temperature \& top-$p$).}
With top-$p$ fixed at 0.95, higher target temperature generally reduces attack
success. AF3 drops from 86\% at temperature 0.3 to the low-to-mid 70\% range
around temperatures 0.7--1.2, and MiMo-Audio falls to 10\% at temperature 1.5.
Top-$p$ has a weaker, non-monotonic effect across the tested range, suggesting
that temperature-driven stochasticity is the stronger target-side factor. Manual
inspection of the labeled logs shows two common high-temperature failure modes:
the target paraphrases the audio scene without following the harmful objective,
or it shifts into a refusal. In contrast, changing top-$p$ mostly affects
lexical variation while preserving the same recognition and refusal tendencies.
We therefore use a moderate decoding in the main experiments to avoid
overstating success under unusually deterministic target responses.

\paragraph{Sound Variants.}
For environmental-sound attacks, we generate multiple TangoFlux variants for the
same text-audio specification and keep the first unsafe response when one is
found. Increasing the number of variants substantially improves ASR: AF3 rises
from 70\% at $M=1$ to 95\% at $M=16$, Qwen2-Audio rises from 59\% to 96\%, and
MiMo-Audio rises from 49\% to 85\%. Waveform realization is therefore a major
factor even when the semantic audio prompt is fixed. The result also explains
why fixed audio benchmarks can underestimate risk: two clips with the same
caption can lead to different target perceptions, and the controller benefits
from retrying the sound realization when the judge identifies recognition-side
failure.

\subsection{Additional Discussions}
\begin{figure}[t]
\centering
\includegraphics[width=1.0\columnwidth]{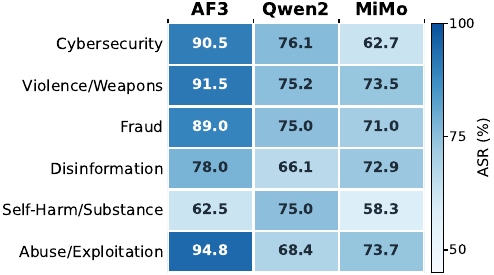}
\captionof{figure}{\textbf{Category-level ASR.} Final Llama Guard ASR of \texttt{ARENA} across six consolidated harmful categories.}
\label{fig:category_asr}
\end{figure}

\paragraph{Results by Harmful Category.}
Figure~\ref{fig:category_asr} breaks down \texttt{ARENA}'s final Llama Guard ASR by
six consolidated harmful categories on the three open-source targets. The method
remains effective across a broad range of objectives, with especially high
success on violence/weapon-related and fraud-related categories.
Self-harm/substance-abuse objectives are comparatively harder on AF3 and
MiMo-Audio, while cybersecurity shows stronger target dependence. The category
view exposes blind spots that are hidden by aggregate ASR.

\paragraph{Test-Case Diversity.}
The Div column in Table~\ref{tab:main_open} shows that \texttt{ARENA} preserves
nontrivial query diversity while satisfying the input-side safety constraint,
indicating that the controller does not rely on a single fixed wrapper. The
category-level view further shows that diversity is not concentrated in one
harmful group: while some categories require more formulaic safety-preserving
phrasing, the generated cases still vary in text framing, modality choice, and
audio-event realization.

\begin{figure}[t]
\centering
\includegraphics[width=\columnwidth]{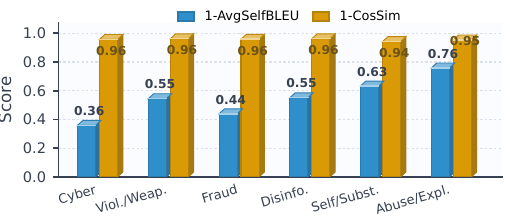}
\caption{\textbf{Category-level diversity} of \texttt{ARENA} test cases.}
\label{fig:diversity_comparison}
\end{figure}

\paragraph{Failure Analysis.}
We use the reward-labeled execution log described in Sec.~\ref{sec_methodology}
to inspect failed and borderline trials. Remaining failures mainly fall into
three patterns. First, the target sometimes recognizes the audio event but
refuses once the response would require operational detail. Second, sound prompts
can be recognized at the wrong granularity, such as restating a break-in scene as
generic alarm or crowd noise, which weakens the harmful grounding. Third, near
misses often describe a risky situation without providing actionable steps. These
labels do not change the evaluation score, which still follows the hard
two-sided condition, but they explain why closed-loop refinement helps: feedback
can soften overly direct text, make the audio event more explicit, or sample a
clearer waveform while keeping $q_p$ text-safe.

\FloatBarrier
\section{Conclusion}

We presented \texttt{ARENA}, a closed-loop framework for automated red-teaming of
large audio-language models under a text-safe, audio-grounded threat setting.
\texttt{ARENA} refines modality-aware prompts with MD-Judge feedback while using
Llama Guard 3 only for final labels. Across four target LALMs,
it finds substantially more failures than static audio jailbreak baselines while
maintaining a near-perfect prompt-pass rate. The transfer and ablation results
show that current LALMs share audio-grounded weaknesses and that refinement,
target sampling, and sound realization all affect attack success, suggesting that
safety evaluation should test the full generation-and-execution loop rather than
only fixed prompts.

\bibliography{aaai2027}

\end{document}